\documentclass[aps,prl,twocolumn,superscriptaddress,10pt,showpacs]{revtex4-1}

\usepackage[utf8x]{inputenc}
\usepackage{graphicx}
\usepackage{amsmath,amssymb}
\usepackage{datetime}
\usepackage{color}
\usepackage[svgnames]{xcolor}
\usepackage{array}
\usepackage[
pdftitle={Floquet dynamics in driven Fermi-Hubbard systems},    
pdfauthor={Michael Messer, Kilian Sandholzer, Frederik Goerg, Joaquin Minguzzi, Remi Desbuquois, Tilman Esslinger},     
pdfsubject={Optical Lattice},   
pdfcreator={Michael Messer, Kilian Sandholzer, Frederik Goerg, Joaquin Minguzzi, Remi Desbuquois, Tilman Esslinger},   
pdfkeywords={Optical Lattice } {Driven Fermi-Hubbard Model } {Ultracold Fermions } {Floquet }, 
colorlinks=True,linkcolor=DarkSlateBlue,citecolor=DarkBlue,urlcolor=DarkBlue,
	pdfstartview=FitH,bookmarks=False,pdfpagemode=UseNone
]{hyperref}

\begin{document}

\title{Floquet dynamics in driven Fermi-Hubbard systems}

\author{Michael Messer}
\author{Kilian Sandholzer}
\author{Frederik G\"org}
\author{Joaqu\'in Minguzzi}
\author{R\'emi Desbuquois}
\author{Tilman Esslinger}
\affiliation{Institute for Quantum Electronics, ETH Zurich, 8093 Zurich, Switzerland}

\date{\today}

\begin{abstract}
We study the dynamics and timescales of a periodically driven Fermi-Hubbard model in a three-dimensional hexagonal lattice. 
The evolution of the Floquet many-body state is analyzed by comparing it to an equivalent implementation in undriven systems. 
The dynamics of double occupancies for the near- and off-resonant driving regime indicate that the effective Hamiltonian picture is valid for several orders of magnitude in modulation time. 
Furthermore, we show that driving a hexagonal lattice compared to a simple cubic lattice allows to modulate the system up to 1~s, corresponding to hundreds of tunneling times, with only minor atom loss. 
Here, driving at a frequency close to the interaction energy does not introduce resonant features to the atom loss.
\end{abstract}

\maketitle

Floquet engineering is a versatile method to implement novel, effectively static Hamiltonians by applying a periodic drive to a quantum system  \cite{Goldman2014, Bukov2015b, Eckardt2017}.
For long timescales, a limitation for this method to create interesting many-body states is eventually the heating to an infinite temperature, caused by the presence of integrability breaking terms such as interactions  \cite{Lazarides2014, DAlessio2014a}. 
For very short time scales, an obvious limit is set by the duration of a single cycle, which cannot be captured by a static Hamiltonian. 
In general, the launch of the drive causes complex dynamics on different timescales in a many-body system \cite{Poletti2011, Weidinger2016, Novicenko2017}.
Theoretical considerations suggest that an effective Hamiltonian picture can still remain valid for some intermediate timescale required to create many-body phases \cite{Abanin2015, Abanin2015a, Bukov2015, Mori2016, Weidinger2016, Canovi2016, Peronaci2017a, Moessner2017a,  Herrmann2018}.
Developing an experimental approach to identify relevant timescales in a periodically driven quantum system with interactions is thus a timely challenge.

\begin{figure}[h!]
    \includegraphics[width=\columnwidth]{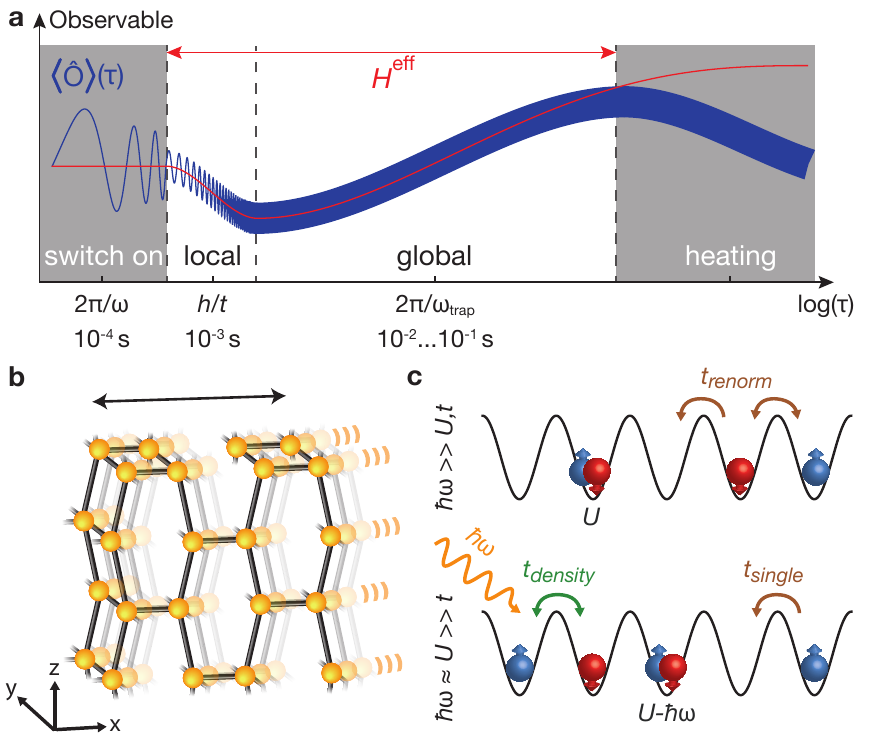}
    \caption{Periodically driven Fermi-Hubbard model. 
    {\bf (a)} Altering the effective Hamiltonian ($H^{\text{eff}}$) affects the underlying many-body state and results in a change of the measured observable within different timescales of the system.  
    A full time evolution of the observable is depicted in blue, while the time evolution under $H^{\text{eff}}$ is plotted in red. 
    Deviations to the expected behavior in the effective Hamiltonian can arise for long modulation times when heating processes dominate. 
    {\bf (b)} Three-dimensional hexagonal structure to realize the driven Fermi-Hubbard model.
    {\bf (c)} Schematics of the tight-binding model of the effective Hamiltonian. 
    For off-resonant driving ($\hbar\omega \gg U,t$) the interactions $U$ are unaffected while the tunneling $t$ is renormalized. 
    In contrast, for near-resonant driving ($\hbar\omega \approx U \gg t$) the system exhibits a reduced effective interaction and a density assisted hopping process which is different from the single particle hopping. 
	}\label{fig1}
\end{figure}

In this Letter, we investigate the Floquet dynamics of a periodically driven Fermi-Hubbard model, which is realized with interacting fermions in a three-dimensional optical lattice. 
Our approach allows us to experimentally compare the evolution of an observable in a driven system with the equivalent dynamics in an undriven Hamiltonian.
The evolution of the entire many-body state is complex (see Fig.~\ref{fig1}a) - while local processes, like the tunneling, play a role on short timescales, the trapping potential sets a timescale for global thermalization.  
In addition, deviations to the expected behavior in the effective Hamiltonian might arise for very long modulation times. 
In the comparison, we analyze this evolution of the many-body state due to a change of (effective) Hubbard parameters and disentangle it from heating in driven systems which cannot be captured by an effective static model.
The latter can be understood as unwanted absorption processes, which in the presence of interactions may be resonant at any driving frequency, since the energy spectrum becomes continuous \cite{Eckardt2015, Bilitewski2015, Bukov2016, Reitter2017}.
Although resonant processes can be desired to realize a specific Floquet Hamiltonian \cite{Parker2013, Aidelsburger2013a, Liberto2014, Bermudez2015, Goldman2015, Mentink2015, Meinert2016, Kitamura2016,  Coulthard2017, Desbuquois2017, Tai2017, Goerg2018, Baum2018, Fujiwara2018}, a general understanding of the dynamics of strongly correlated driven quantum states over several orders of magnitude in evolution time remains challenging \cite{Genske2015, Lellouch2016, Wang2017, Keles2017, Herrmann2018, Qin2018}.

For our measurements we prepare a degenerate fermionic cloud with $N=38(4) \times 10^3$ interacting, ultracold $^{40}\text{K}$ atoms equally populating two magnetic sublevels of the $F=9/2$ hyperfine manifold at a temperature of 10(1) \% of the Fermi temperature. 
The atoms are then loaded into the lowest band of a three-dimensional optical lattice with hexagonal geometry \cite{Tarruell2012}. 
The hexagonal lattice in the xz-plane is a bipartite lattice with sublattices $\mathcal{A}$ and $\mathcal{B}$ and is stacked along the y-direction (see Fig.~\ref{fig1}b and \cite{Supplementary}).
The position of the retro-reflecting lattice mirror along the x-direction is then periodically modulated using a piezo-electric actuator at frequency $\omega/(2\pi)$ and amplitude $A$. 
To compare the evolution of the many-body state under the driven and undriven Hamiltonian we measure the fraction of doubly occupied sites $\mathcal{D}$ for different times. 
This probes the many-body state for a given set of tunneling, interactions, and atom number \cite{Jordens2010, Uehlinger2013}.

For an off-resonant modulation, where the driving frequency is the dominant energy scale ($\hbar \omega \gg U,t$) our system is described by the effective Hamiltonian \cite{Eckardt2005a, Lignier2007, Zenesini2009, Goerg2018}:
\begin{align}
\hat{H}^{\text{eff}}_{\text{off-res}}  
&= -  t_{x}  \mathcal{J}_0(K_0) \sum_{\langle \text{i,j} \rangle_x , \sigma} \hat{c}^{\dagger}_{\text{i} \sigma}\hat{c}_{\text{j} \sigma} -t_{y,z} \sum_{\langle \text{i,j} \rangle_{y,z} , \sigma} \hat{c}^{\dagger}_{\text{i} \sigma}\hat{c}_{\text{j} \sigma} \nonumber \\ & + U \sum_{\text{i}} \hat{n}_{\text{i} \downarrow}\hat{n}_{\text{i} \uparrow} + \sum_{\text{i}} V_{\text{i}} \hat{n}_{\text{i}}\, ,
\label{Heff_off_res}
\end{align}
where $\hat{c}^{\dagger}_{\text{i} \sigma}$ ($\hat{c}_{\text{i} \sigma}$) are the creation (annihilation) operators of one fermion with spin $\sigma=\uparrow,\downarrow$ at lattice site $\text{i}$ and $\hat{n}_{\text{i} \sigma}=\hat{c}^{\dagger}_{\text{i} \sigma}\hat{c}_{\text{i} \sigma}$.
The tunneling rates $t_{x,y,z}$ connect nearest neighbors $\langle \text{i,j} \rangle$ along $x,y,z$ and $U$ is the on-site interaction energy.  
The last term represents the harmonic confinement of the trap, characterized by the mean trapping frequency $\overline{\omega}$ \cite{Supplementary}. 
In the off-resonant regime the tunneling energy along the driving direction ($x$) is renormalized by the zeroth-order Bessel function $\mathcal{J}_0$ with the dimensionless driving amplitude $K_0=mA\omega d_x/\hbar$ in the argument, where $m$ is the mass and $d_x$ the lattice spacing \cite{BrickWall}.

When we modulate near-resonantly to the interaction energy ($\hbar \omega \approx U \gg t$) the effective Hamiltonian is to lowest order in $1/\omega$ given by  \cite{Bermudez2015, Itin2015, Bukov2016b, Goerg2018}: 
\begin{align}
\hat{H}^{\text{eff}}_{\text{res}}  
&= - \sum_{\langle \text{i,j}\rangle_x , \sigma} \left( t_{x}^{\text{eff,} 0} \hat{g}_{\text{ij}\overline{\sigma}} + t_{x}^{\text{eff,}\text{D}} \left[ \hat{h}^{\dagger}_{\text{ij}\overline{\sigma}} + \text{h.c.} \right] \right)  
 \hat{c}^{\dagger}_{\text{i} \sigma}\hat{c}_{\text{j} \sigma} -t_{y,z} \nonumber \\ \label{Heff_res}
&\sum_{\langle \text{i,j} \rangle_{y,z} , \sigma} \hat{c}^{\dagger}_{\text{i} \sigma}\hat{c}_{\text{j} \sigma}  
+ \left(U - \hbar \omega \right) \sum_{\text{i}} \hat{n}_{\text{i}\downarrow}\hat{n}_{\text{i} \uparrow} + \sum_{\text{i}} V_{\text{i}} \hat{n}_{\text{i}} \, .
\end{align}
Here, the interaction is effectively modified to a value $U^{\text{eff}}=U - \hbar \omega$. This can be understood as exchange of photons with the drive. 
In addition, we have to differentiate between tunneling events which keep the number of double occupancies constant $\left[ t_{x}^{\text{eff,} 0} \, \text{with} \, \hat{g}_{ij \overline{\sigma}}=(1-\hat{n}_{i \overline{\sigma}})(1-\hat{n}_{j \overline{\sigma}})+\hat{n}_{i \overline{\sigma}}\hat{n}_{j \overline{\sigma}}  \, \text{and}\, \overline{\uparrow} = \downarrow \right]$ and those which increase or decrease it by one unit $[ t_{x}^{\text{eff,}\text{D}} \, \text{with} \, \hat{h}^{\dagger}_{ij \overline{\sigma}}=\pm \hat{n}_{i \overline{\sigma}}(1-\hat{n}_{j \overline{\sigma}})$, where the positive sign is valid for $i<j$ and vice versa$]$. 
The tunneling of a particle to an empty neighboring site is unaffected by the interaction resonance and we obtain $t_{x}^{\text{eff,} 0} = t_{x}  \mathcal{J}_0(K_0) $, as in the off-resonant case.
In contrast, if a double occupancy is involved in the tunneling process we get $t_{x}^{\text{eff,}\text{D}} = t_{x}  \mathcal{J}_1(K_0)$, thereby realizing density assisted tunneling processes \cite{Ma2011, Chen2011, Liberto2014, Mentink2015, Meinert2016, Kitamura2016, Coulthard2017, Desbuquois2017, Goerg2018}.
Fig.~\ref{fig1}c presents a schematic overview of the microscopic processes for the off- and near-resonant drive.

\begin{figure}
    \includegraphics{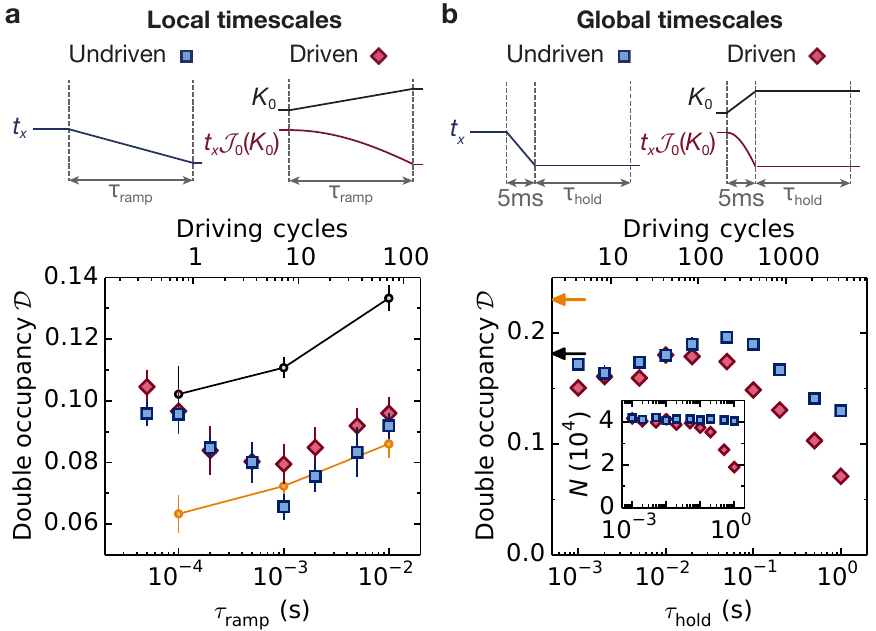}
    \caption{Off-resonantly modulated Fermi-Hubbard model.  
    {\bf (a)} Starting from a $t_{x,y,z}/h=(200(30),40(3),40(3))\, \text{Hz}$ hexagonal lattice we either increase the driving amplitude to $K_0=1.69(2)$ or perform a lattice ramp (undriven lattice) to imitate the same final $t_x=t_x^{\text{eff}}/h=80(10) \, \text{Hz}$.
    The evolution of double occupancy fraction $\mathcal{D}$ on a local timescale for the driven (red diamonds) and undriven lattice (blue squares) as a function of the ramp-up time $\tau_{\text{ramp}}$ at $U/h = 500(30)\, \text{Hz}$. 
	Reference values when loading directly into the starting (final) lattice are shown in black (orange) circles. 
{\bf (b)} 
For the global timescales we start with a $t_{x,y,z}/h=(510(100),100(6),100(10))\, \text{Hz}$ lattice and ramp up the drive within 5~ms or directly decrease $t_x$ to $t_{x}^{\text{eff}}/h=210(40)\, \text{Hz}$. 
Measured $\mathcal{D}$ at $U/h = 700(20)\, \text{Hz}$ as a function of the modulation time $\tau_{\text{hold}}$ at constant $K_0=1.69(3)$ (red diamonds) and its counterpart in the undriven case (blue squares). 
    Arrows indicating the reference values in the starting (final) lattice are shown in black (orange).
    The inset shows the corresponding number of atoms $N$ as a function of $\tau_{\text{hold}}$. 
    Data points in {\bf a} ({\bf b}) are the mean and standard error of 5 (10) individual measurements at different times within one driving period (see \cite{Supplementary}).
	}\label{fig2}
\end{figure}

In a first set of measurements we compare the evolution of the fraction of doubly occupied sites ($\mathcal{D}=2/N \sum_{\text{i}} \langle \hat{n}_{\text{i} \downarrow}\hat{n}_{\text{i} \uparrow} \rangle$) under a change of the Hamiltonian for off-resonant modulation and an equivalent ramp in the undriven system.  
We linearly ramp up the amplitude $K_0=1.69(2)$ at a modulation frequency $\omega/(2\pi) = 7.25 \, \text{kHz}$  within a variable ramp time $\tau_{\text{ramp}}$ (see Fig.~\ref{fig2}a). 
After the modulation ramp is completed, the tunneling in $x$ is reduced to $t_x^{\text{eff}}= t_x \mathcal{J}_0(K_0) \approx 0.4 t_x$.
We achieve an equivalent change of $t_x$ by ramping the lattice depth of an undriven system. 
Interestingly, the measured $\mathcal{D}$ in the driven system follows the results of the undriven lattice for all timescales.
Both data sets start at the reference of the initial lattice and reach the reference of the final lattice within 1~ms.  
We can explain this effect with a local change of the population of double occupancies and single particles due to an increased $U/t_x^{\text{eff}}$.
Already a single driving cycle reduces the level of $\mathcal{D}$, indicating the effective Hamiltonian picture can be valid on such short timescales.

To focus on the global timescales we use a lattice with faster tunneling and first ramp up the driving within 5~ms, which we have observed is adiabatic with respect to local timescales \cite{Supplementary}.
At maximal amplitude of $K_0 = 1.69(3)$ and driving frequency of 4.25~kHz we vary the modulation time $\tau_{\text{hold}}$ and compare the resulting change of $\mathcal{D}$ with a ramp in the undriven lattice (see Fig.~\ref{fig2}b).
We observe a slowly increasing $\mathcal{D}$ and both measurements follow each other up to $\tau_{\text{hold}}=50\, \text{ms}$, which corresponds to more than 200 driving cycles.
As a result, even at timescales where the trap redistribution plays a role, the off-resonantly modulated Fermi-Hubbard model is captured by the effective Hamiltonian in Eq.~\ref{Heff_off_res}. 
For $\tau_{\text{hold}}>\, 0.1 \text{s}$ we observe a decrease of $\mathcal{D}$, even in the undriven case, which we attribute to technical heating for a trapped system at intermediate interactions \cite{Jordens2010}. 
In both cases, this heating prevents a full redistribution in the trap as the adiabatic reference value is not fully reached (orange arrow). 
On a similar timescale, the driven lattice exhibits a loss of atoms (see inset of Fig.~\ref{fig2}b) which will be analyzed in more detail in Fig.~\ref{fig4}.

\begin{figure}[bt]
    \includegraphics{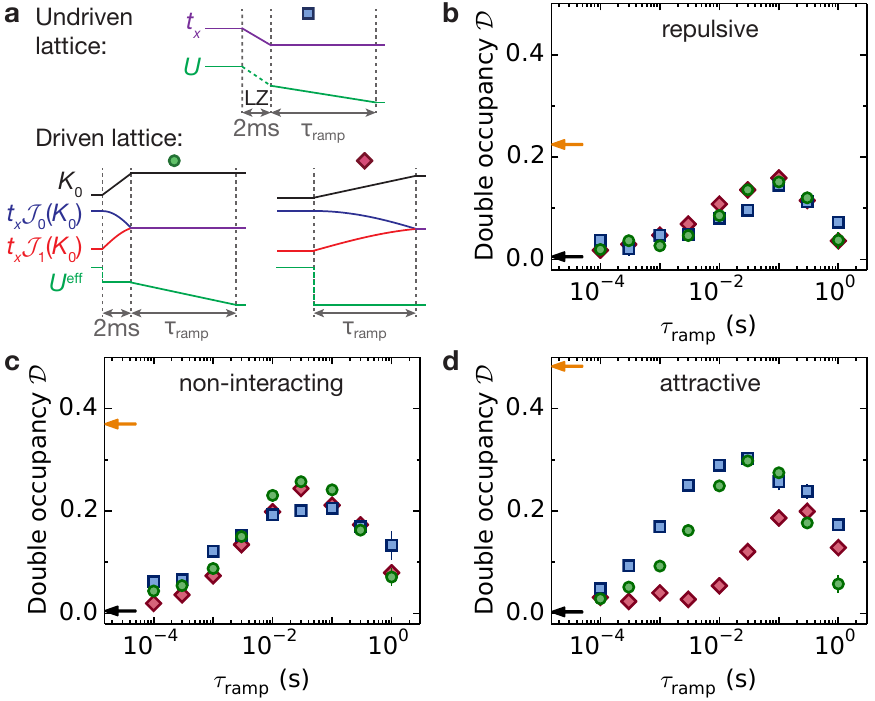}
    \caption{Near-resonantly modulated Fermi-Hubbard model. 
    {\bf (a)} In the undriven lattice (blue squares) we change $U$ in 2~ms using a Landau-Zener (LZ) transfer to a different spin state and also lower $t_x$. 
    For near-resonant driving we either increase $K_0$ to 1.43(2) on a fixed timescale (2~ms) at $U_{\text{load}}=4.63(10) \, \text{kHz}$ and then tune the interactions in the driven system within a variable time $\tau_{\text{ramp}}$ (green circles) or we first prepare the system at the final interaction $U$ and then vary $\tau_{\text{ramp}}$ of the $K_0$ ramp (red diamonds). 
    {\bf (b-d)} Dynamics of $\mathcal{D}$ for different $\tau_{\text{ramp}}$ and three values of the (final) effective interaction $U^{\text{eff}}/h=[0.69(8), -0.02(6), -0.72(5)] \, \text{kHz}$ or the corresponding static counterpart $U^{\text{stat}}$. 
    Arrows indicate the reference values when adiabatically loading the atoms in the starting lattice $t_{x,y,z}/h=(200(30),100(10),100(10))\, \text{Hz}$ (black) or the final lattice with reduced tunneling $t_{x}^{\text{eff}}/h=110(20)\, \text{Hz}$ and $U^{\text{eff}}$ (orange). 
    Data points are the mean and standard error of 5 individual measurements at different times within one driving period \cite{Supplementary}.
	}\label{fig3}
\end{figure}

\begin{figure}[bt]
    \includegraphics{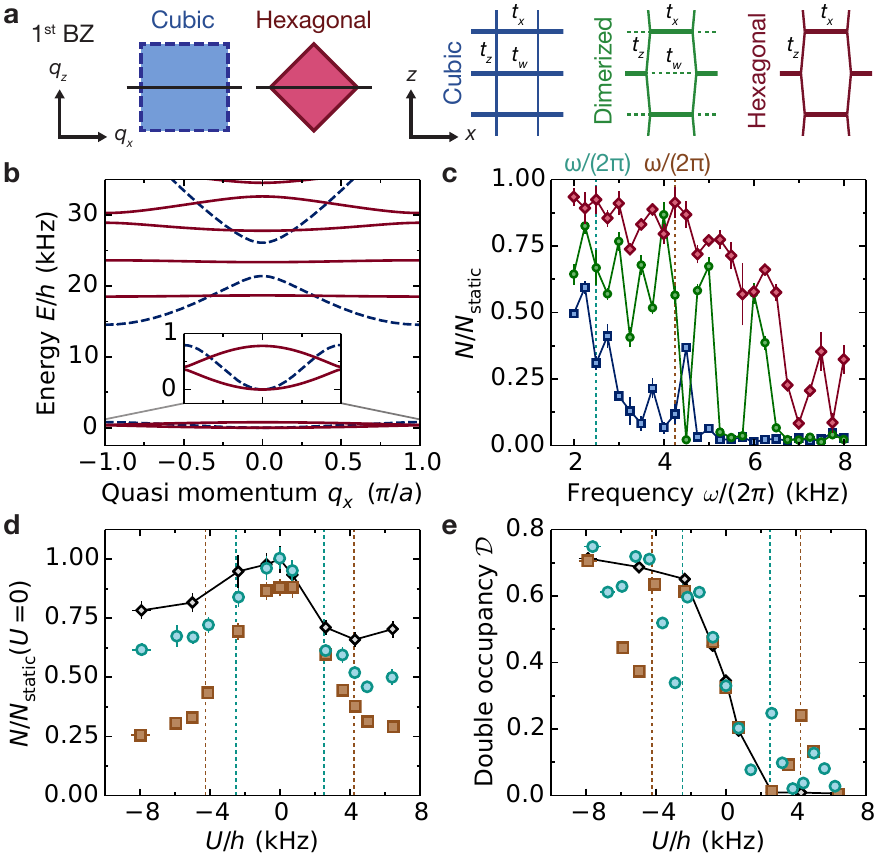}
    \caption{Minimizing atom loss in driven 3D optical lattices. 
    {\bf (a)} Schematics of the first Brillouin zone ($1^{\text{st}}$~BZ) of the simple cubic (SC, blue) and hexagonal (HC, red) lattice and their real space geometry including an intermediate dimerized (D, green) lattice.
    All three lattice configurations have $t_x/h\approx200 \, \text{Hz}$ but a different $t_w/h$ (SC: 200(10)~Hz, D: 36(2)~Hz and HC: $< 1\, \text{Hz}$).
    {\bf (b)} Calculated band structure ($q_z=0$) for the SC (HC) lattice plotted in blue dashed (red). 
    The HC lattice allows to tune the bandgap and dispersion of higher band while the bandwidth of the lowest band remains similar compared to the SC lattice (see inset). 
    {\bf (c)} Remaining fraction of atoms $N/N_{\text{static}}$ when driving for 1~s at fixed $K_0=1.43(2)$ for different frequencies $\omega/(2\pi)$ at $U/h=710(20)\,\text{Hz}$, where $N_{\text{static}}$ is the remaining atom number after holding the system for 1~s in the undriven lattice. 
    The two vertical dashed lines indicate the driving frequencies used in {\bf d,e}.
    {\bf (d)} Normalized atom number $N/N_{\text{static}}(U=0)$ in the HC lattice after a modulation of 1~s at  $K_0=1.43(2)$ for two different frequencies $\omega/(2\pi)=4.25 \, \text{kHz}$ (orange) and $\omega/(2\pi)=2.5 \, \text{kHz}$ (cyan) and without drive (black) as a function of $U$.
    {\bf (e)} $\mathcal{D}$ corresponding to the measurements in {\bf (d)} but for a modulation time of 5~ms.
    Dashed vertical lines in {\bf d,e} correspond to the resonance condition $U = \pm \hbar \omega$. 
    Data points in {\bf c} ({\bf d,e}) are the mean and standard error of 3 (5) individual measurements at different times within one driving period.
    Error bars in $U$ result from the uncertainty of the calibration. 
	}\label{fig4}
\end{figure}

In a second set of measurements we probe the validity of the effective Hamiltonian for a near-resonant modulation ($\hbar \omega \approx U \gg t$).
In contrast to the measurements so far, we prepare our initial system in a Mott insulating state ($U/h>4.6(1) \, \text{kHz}$) with negligible $\mathcal{D}$ and follow two different driving protocols with $\omega/(2\pi)=3.5 \, \text{kHz}$ (see Fig.~\ref{fig3}a). 
We choose $K_0=1.43(2)$ such that tunneling is independent of the density ($\mathcal{J}_0(K_0)=\mathcal{J}_1(K_0)$, see Eq. \ref{Heff_res}) which allows us to compare the system to an undriven parameter ramp.
We either switch on $K_0$ to its maximal value within a variable time $\tau_{\text{ramp}}$ at the final interaction $U$ (red diamonds) or we follow a more intricate protocol. 
For this, we first ramp up $K_0$ in 2~ms at $U_{\text{load}}$ detuned from resonance and then adjust $U_{\text{load}} \rightarrow U$ while modulating \cite{Supplementary}. 
For near-resonant driving, measurements on isolated double wells have shown that states form avoided crossings when coupled resonantly, resulting in a ramp dependent control of Floquet states \cite{Desbuquois2017}.

For comparison, in an undriven lattice we need to ramp $t_x$ to mimic the renormalized tunneling and additionally reduce the initial interactions to $U -\hbar \omega$ (see Fig.~\ref{fig3}a). 
The latter is achieved by using the Feshbach resonances of two different spin mixtures, which have different values of $U$ at the given constant magnetic field. 
Hence, we perform a Landau-Zener transfer between two internal spin states, thereby reducing the initial strong repulsive interactions to weakly repulsive values within 2~ms \cite{Supplementary}.
In a final step, we ramp the magnetic field on a variable time to reach the final value corresponding to $U^{\mathrm{eff}}$.

The dynamics of $\mathcal{D}$ as a function of $\tau_{\text{ramp}}$ are shown in Fig.~\ref{fig3}.
We choose three different detunings $U -\hbar \omega$ which result in a weakly repulsive, non-interacting for a modulation on resonance, and weakly attractive effective interaction.
In all measurements $\mathcal{D}$ steadily increases since the initial Mott insulating regime is altered by the reduced effective interactions, but does not reach the reference value when adiabatically loading a weakly interacting cloud. 
Both the weakly repulsive, as well as the non-interacting case, do not show a difference in $\mathcal{D}$ for the two modulation protocols and follow the expectation of the undriven system.

However, for effective attractive interactions we observe a strong deviation.  
Here, the system clearly has a memory of the ramping protocol although all schemes reach the same final Hubbard parameters. 
By ramping up $K_0$ away from resonance and then tuning $U$ in the driven system (green circles), $\mathcal{D}$ gets closer to the undriven reference (blue squares). 
This driving protocol is thus suitable to realize a many-body state with effective attractive interactions. 
In contrast, the level of $\mathcal{D}$ is equivalent for $U^{\text{eff}}/h= \pm 700 \, \text{Hz}$ for the other driving scheme (red diamonds). 
Similar to the off-resonant case, we observe a decrease in $\mathcal{D}$ for long timescales ($\tau_{\text{ramp}}> 0.1\, \text{s}$), even in the undriven lattice. 
Since all three schemes show a similar loss of $\mathcal{D}$, heating seems to be unrelated to the drive. 
In addition, on long timescales we observe atom loss in the driven system.

We have seen that deviations from the undriven system arise when the modulation leads to additional atom loss. 
In our measurements these losses have been minimized by a smart choice of geometry, namely a hexagonal lattice with tunable bandgaps. 
In general, for a non-interacting system, atom loss can be caused by resonant coupling to energetically higher bands in single or multiphoton processes \cite{Weinberg2015, Jotzu2015, Strater2016, Quelle2017, Sun2018, Flaschner2018}. 
As a result, larger bandgaps and less dispersive higher bands broaden the frequency window suitable for a Floquet system and reduce the atom loss \cite{Sun2018}. 
By using an anisotropic lattice along the modulation direction ($t_x \neq t_w$) we tune the bandgap and dispersion of higher bands, while the bandwidth of the lowest band is kept on a similar level (see Fig.~\ref{fig4}a,b).
Fig.~\ref{fig4}c shows the remaining number of atoms after modulating for 1~s at various frequencies with $K_0=1.43(2)$. 
We compare the loss for the hexagonal lattice used in the measurements of the evolution of $\mathcal{D}$ with a dimerized and simple cubic lattice in the weakly repulsive regime ($U/h=0.71(2) \, \text{kHz}$).
While the atom loss in the simple cubic lattice is quite severe, a dimerization significantly improves the situation and a minimal loss rate is reached for the hexagonal lattice \cite{Shaking}.

To further investigate the role of interactions in the hexagonal lattice we measure the atom loss at different $U$ when modulating for 1~s ($K_0=1.43(2)$)  (see Fig.~\ref{fig4}(d)). 
We compare this data with measurements where the atoms are held in a static lattice for 1~s. 
In general, atom loss is increased for stronger $U$, both in the static and driven hexagonal lattice. 
However, interactions do not introduce new resonant features even though the fraction of double occupancies $\mathcal{D}$ shows the expected reduction (increase) when driving resonantly to the attractive (repulsive) interactions (see Fig.~\ref{fig4}(e)).

In conclusion, we have demonstrated the validity of the effective Hamiltonian over several orders of magnitude in evolution time for near- and off-resonant modulation. 
Furthermore, our results show that the driven Fermi-Hubbard model can be implemented on realistic experimental timescales, since atom loss and technical heating dominate only after relatively long modulation times. 
In future work, a direct comparison to theoretical simulations can provide further understanding of driven interacting systems and allows us to investigate Floquet prethermal states \cite{Abanin2015a, Bukov2015, Bukov2016, Weidinger2016, Canovi2016, Herrmann2018}.
Moreover, a successful implementation and benchmarking of driven many-body states opens the possibility to investigate the $t-J$ model \cite{Coulthard2018} and correlated hopping systems \cite{Rapp2012a, Liberto2014, Meinert2016, Goerg2018}.
In addition, we can extend our technique to complex density dependent tunneling to realize exotic interacting topological systems and dynamical gauge fields \cite{Keilmann2011, Goldman2013, Greschner2014b, Bermudez2015}.

\vspace{1mm}

\begin{acknowledgments}
We thank H. Aoki, J. Coulthard, D. Jaksch, Y. Murakami, and P. Werner for insightful discussions and G. Jotzu for careful reading of the manuscript. We acknowledge SNF (Project Number 200020\_169320 and NCCR-QSIT), Swiss State Secretary for Education, Research and Innovation Contract No. 15.0019 (QUIC) and ERC advanced grant TransQ (Project Number 742579) for funding.
\end{acknowledgments}

\clearpage

\makeatletter
\setcounter{section}{0}
\setcounter{subsection}{0}
\setcounter{figure}{0}
\setcounter{equation}{0}
\renewcommand{\bibnumfmt}[1]{[S#1]}
\renewcommand{\thefigure}{S\@arabic\c@figure}
\renewcommand{\theequation}{S\@arabic\c@equation}
\makeatother

\centerline{\Large \textbf{Supplemental material}}

\subsection{General preparation}

The experiment starts with a gas of $^{40}$K fermionic atoms in the two magnetic sublevels $m_F=-9/2,-7/2$ of the $F=9/2$ manifold, which is trapped in a harmonic optical dipole trap. The atoms are evaporatively cooled down to quantum degeneracy at a scattering length $a = 116(1)$ $a_0$ ($a_0$ is the Bohr radius) and we prepare a spin-balanced cloud of $38(4)\times 10^3$ atoms at a temperature $T=0.10(1)/T_F$ ($T_F$ is the Fermi temperature). For attractive and weakly repulsive interactions we use a $-9/2,-7/2$ mixture and for strongly repulsive a $-9/2,-5/2$ one. The interactions are tuned with Feshbach resonances around 202.1 G and 224.2 G for $-9/2,-7/2$ and $-9/2,-5/2$ mixtures, respectively. The latter is prepared with a Landau-Zener transfer that flips the $-7/2$ spin component into the $-5/2$ one.

The three-dimensional optical lattice is made out of four retro-reflected laser beams of wavelength $\lambda = 1064 \, \text{nm}$. The lattice potential seen by the atoms is
\begin{eqnarray} V(x,y,z) & = & -V_{\overline{X}}\cos^2(k
x+\theta/2)-V_{X} \cos^2(k
x)\nonumber\\
&&-V_{\widetilde{Y}} \cos^2(k y) -V_{Z} \cos^2(k z) \nonumber\\
&&-2\alpha \sqrt{V_{X}V_{Z}}\cos(k x)\cos(kz)\cos\varphi , 
\label{Lattice}
\end{eqnarray}
with $k=2\pi/\lambda$ and $x,y,z$ are the three experimental axes. The lattice depths $V_{\overline{X},X,\tilde{Y},Z}$ are measured in units of the recoil energy $E_R=h^2/2m\lambda^2$ ($h$ is the Planck constant and $m$ the mass of the atoms) and each of them is individually calibrated using amplitude modulation on a $^{87}$Rb Bose-Einstein condensate. The visibility $\alpha=0.99(1)$ is also calibrated using amplitude modulation on a $^{87}$Rb Bose-Einstein condensate, but in an interfering lattice configuration. The phases $\theta$ that fixes the geometry of the lattice is stabilized to $\theta=1.000(2)\pi$.
The Hubbard parameters $t$ and $U$ are numerically calculated from the Wannier functions of the lattice potential, which we obtain from band-projected position operators \cite{Uehlinger2013}.
The bandwidth $W$ of the single band tight-binding model is defined as $W=2\sum_i t_i$, where $i$ sums all nearest-neighbor tunneling rates $t_i$ of the lattice geometry. 

\subsection{Periodic driving}

The periodic driving is implemented with a piezo-electric actuator that modulates the position of the retro-mirror for the $X$ and $\overline{X}$ lattice beams at a frequency $\omega/2\pi$ and displacement amplitude $A$. This shifts the phase of the retro-reflected $X$ and $\overline{X}$ lattice beams with respect to the incoming ones such that the time-modulated ($\tau$) lattice potential is $V(x,y,z,\tau)=V(x-A\cos(\omega\tau),y,z)$. 
The amplitude $A$ is related to the normalized amplitude by $K_0=mA\omega d_x/\hbar$, where $d_x$ is the distance between two sites along the $x$-direction ($\hbar=h/2\pi$). 
The distance $d_x$ changes for different lattice configurations, and any deviations from $\lambda/2$ are included when we estimate $K_0$. 
The exact value of $d_x$ is numerically computed as the distance between the location of the Wannier function on the left and right side of the corresponding lattice bond. 
The values for each lattice configuration are shown in Tables \ref{Table_offRes} to \ref{Table_bandgap}. 
Since the hexagonal geometry is not an ideal brick configuration the driving also slightly modifies the bonds along the $z$-direction. 
Again, the strength is given by $K_0=mA\omega d_x^{\text{vert}}/\hbar$ with $d_x^{\text{vert}}$ as the projected length of the $z$-bonds along the driving direction which can be rewritten as $d_x^{\text{vert}}=\lambda/2-d_x$. 
The modulation amplitude is therefore drastically reduced along the $z$ bonds and only minor renormalization to the tunneling along $z$ occur.

Furthermore, the phase $\varphi=0.0(1)\pi$ is stabilized by periodically modulating the phase of the incoming $X$ and $Z$ lattice beams at the same frequency as the drive using acousto-optical modulators to keep the lattice geometry fixed. 
As the compensation is not perfect, the piezo modulation leads to a residual periodic reduction in the amplitude of the two interfering $X$ and $Z$ beams  by at most 2\%. 
As a result, the tunneling energy $t_x$ is modulated at twice the driving frequency with an amplitude of $\delta t = 0.025t_x$ and its mean is reduced by roughly 2.5\%. 
The effect of this additional amplitude modulation is negligible since the effective driving strength is proportional to $\delta t/(\hbar \omega)$. 
In addition, we use the amplitude of the phase modulation without compensation to calibrate the phase and amplitude of the mirror displacement caused by the piezo-electric actuator.

\subsection{Detection methods}

The detection of double occupancies starts with freezing the dynamics by ramping up the lattice depths to $V_{\overline{X},X,\tilde{Y},Z}=[30,0,40,30] \ E_R$ within $100 \ \mu$s. 
Depending on the exact driving frequency this freeze is partly averaging over the micromotion until the evolution of the system is completely stopped. 
To be insensitive on the micromotion in our experiments we perform individual measurements at different times within one driving period by subsequently freezing the lattice at slightly different modulation times. 
In the deep $V_{\overline{X},X,\tilde{Y},Z}=[30,0,40,30] \ E_R$ lattice we linearly ramp off the periodic driving within 10 ms. 
Then, we use an interaction-dependent radio-frequency transfer that selectively flips the $m_F=-7/2$ atoms on doubly occupied sites to the initially unpopulated $m_F=-5/2$ spin state (and vice versa, depending on whether we start with a $-9/2,-7/2$ or $-9/2,-5/2$ mixture). 
After this, we perform a Stern-Gerlach type scheme by switching off the optical lattice and dipole trap and switching on a magnetic field gradient within 20 ms. After 10 ms of ballistic expansion each of the $-9/2,-7/2,-5/2$ spin components are spatially resolved with an absorption image. For each spin component its spatial density profile is fitted with a Gaussian distribution to estimate the number of atoms on each spin state determining the fraction of double occupancies $\mathcal{D}$.

\begin{figure*}[tb]
    \includegraphics[]{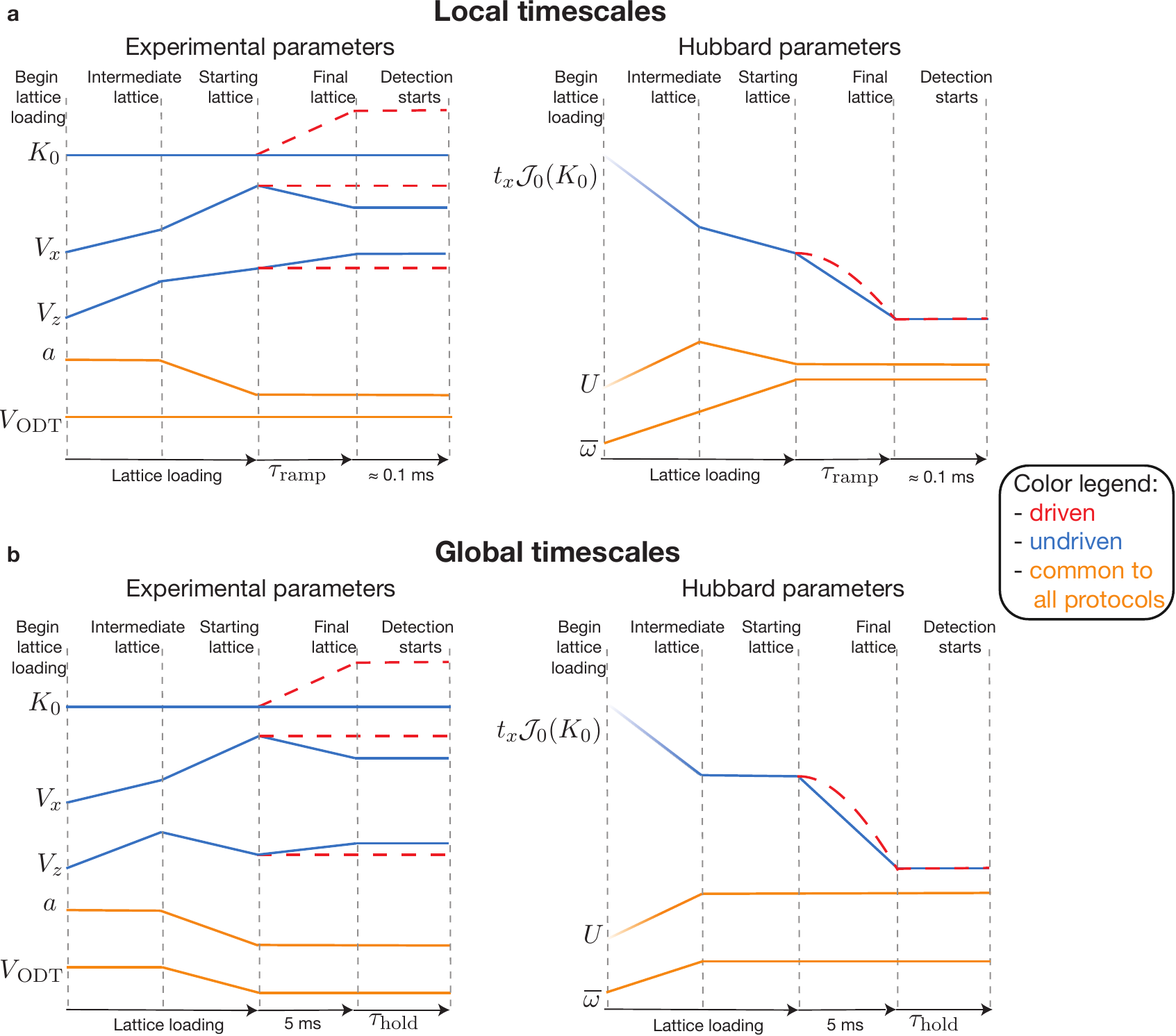}
    \caption{Ramp protocols of the experimental and Hubbard parameters for the off-resonantly driven Fermi-Hubbard model used to study the local {\bf (a)} or global {\bf (b)} timescales. The driving amplitude $K_0$, the lattice depth $V_{x(z)}$, the scattering length $a$, and the depth of the optical dipole trap ($V_{\text{ODT}}$) are shown on the left panel. 
    The Hubbard parameters (tunneling along the shaking direction $t_x$, the on-site interaction $U$ and the mean trap frequency $\overline{\omega}$) are shown on the right panel.   
    A two step protocol via the intermediate lattice allows for a controlled preparation of the many-body state in the deep hexagonal lattice (starting lattice). 
    We either ramp up $K_0$ in the starting lattice of the driven system, or in contrast, we ramp the lattice depths for the undriven lattice to match the same tight binding parameters at the final lattice configuration. 
	}\label{SI_fig_OR_protocol}
\end{figure*}

\subsection{Off-resonant modulation}

The experimental parameters vary slightly between the local and global timescale measurements in the off-resonantly modulated system. 
In the following we present the general preparation scheme, while the actual values for the experimental parameters are given in Table~\ref{Table_offRes}. 
In both cases, we perform multiple steps in order to prepare a many-body state in the hexagonal lattice.

\subsubsection{Local timescales}

We first ramp up all lattice beams within 200 ms to a dimerized lattice configuration with a remaining tunneling link $t_w$ across the hexagonal unit cell (see column \textit{intermediate lattice}).
During an additional 10 ms we then ramp to a hexagonal lattice configuration with negligible tunneling $t_w$. 
This lattice is used as a starting point for the measurement protocol and is referred to as the \textit{starting lattice}.
The system is prepared at an on-site interaction $U/h=500(30)\, \text{Hz} \, (U/W=0.7(1))$ which corresponds to a regime in the crossover from a metal to a Mott insulator. 

We then compare the evolution of $\mathcal{D}$ under a change of Hubbard parameters resulting from the drive or an analogous ramp in the undriven case (see Fig.~\ref{SI_fig_OR_protocol}).  
We either ramp up the amplitude of the drive to $K_0=1.69(2)$ on a varying time $\tau_{\text{ramp}} = \left[ 50 \, \mu \text{s} ,... , 10\, \text{ms} \right]$ or directly perform a third lattice ramp in the undriven system. 
The increase of $K_0$ leads to a renormalization of the tunneling which reaches its minimal values at the end of the ramp. 
The corresponding calculated effective tunneling rates $t_{x,z}^{\text{eff}}$ are given in the \textit{final lattice} column of Table ~\ref{Table_offRes}.
The same change of tunneling is achieved by changing the intensity of the interfering lattice beams $V_X$ and $V_Z$ in the undriven case (see parameters of the undriven system in Table ~\ref{Table_offRes}).
To keep $U$ on a similar level in the intermediate as well as final lattice configuration we ramp the scattering length $a$ to compensate the change of Wannier functions as the lattice parameters are changed (see Fig.~\ref{SI_fig_OR_protocol}).
Additionally, we measure a reference value of $\mathcal{D}$ in a static configuration.
For this, we directly load the starting (final) lattice (see $V_{\overline{X},X,\widetilde{Y},Z}$ of the \textit{starting (final)} lattice in the undriven case) within 10~ms via the intermediate lattice configuration and measure $\mathcal{D}$ for a variable hold time $\tau_{\text{hold}}$.

Due to the additional harmonic confinement of the lattice beams the mean trap frequency increases from $\overline{\omega}_{\text{ODT}}/(2\pi)=54.7(6) \, \text{Hz}$ in the bare optical dipole trap (ODT) to $\overline{\omega}/(2\pi)=85.0(5)\, \text{Hz}$ in the intermediate lattice.
The mean trap frequency is further increased to $\overline{\omega}_{\text{ODT}}/(2\pi)=93.3(8) \, \text{Hz}$ when loading into the starting lattice configuration as the lattice depths are increased. 
This change of trapping frequency might be the reason why the static reference values of $\mathcal{D}$ change as a function of the hold time.

\subsubsection{Global timescales}

The preparation of the many-body state for the global timescales follows a similar scheme (see Fig.~\ref{SI_fig_OR_protocol}). 
However, we use a lattice configuration with larger tunneling to allow for a faster dynamics (see tunneling rates in Table~\ref{Table_offRes}). 
In contrast to the preparation for the local timescales, we remove the additional tunneling link $t_w$ while all other tunnelings remain on a similar level ($t_{x,y,z}/h =(500(90),100(10),100(10))\, Hz$) when ramping from the intermediate to the starting lattice. 
Here, the distance $d_x$ is different from the one used to study local timescales, as the lattice potentials are different. 
As a result, we have to change the amplitude $A$ of the piezo movement in order to reach the same dimensionless driving strength $K_0$.
At the starting lattice we either ramp up the driving amplitude to $K_0=1.69(3)$ or perform an equivalent lattice ramp within 5~ms. 
With this, the effective tunnelings are matched in both protocols. 
Subsequently, we vary the hold time $\tau_{\text{hold}}$ from 1~ms to 1~s in this final lattice configuration for many orders of magnitude and detect $\mathcal{D}$. 
The system is prepared at an interaction of $U/h=710(20)\, \text{Hz} \, (U/W=0.38(6))$, which remains on a similar level from the intermediate to final lattice by ramping down the scattering length.

In contrast to the scheme for the local timescales, we decrease the intensity of the ODT $V_{\text{ODT}}$ when ramping from the intermediate to the starting lattice, to keep $\overline{\omega}/(2\pi)=85.2(8) \, \text{Hz}$ fixed during the full evolution.
We also perform two sets of reference measurements in a static lattice.  
For this, we load during 200~ms an intermediate lattice configuration with the same tunneling $t_{x,y,z}$ as the starting or final lattice configuration but an additional tunneling link $t_w/h=38(2) \, \text{Hz}$.
Within another 10~ms we remove this tunneling $t_w/h<1\,\text{Hz}$ and load either the starting lattice configuration $t_{x,y,z}/h=510(90), 100(6), 102(9), \, \text{Hz}$ or the final lattice configuration $t_{x,y,z}/h=200(30), 100(6), 94(7), \, \text{Hz}$. 
After a hold time of 5~ms we detect $\mathcal{D}$ which we plot as the two reference values in Fig.~\ref{fig2}b.
For the reference measurements, we use the same $U$ and $\overline{\omega}$ as for the driven and undriven comparison.

\begin{table*}[bt]
\setlength\extrarowheight{4pt}
\begin{tabular}{|c || c | c | c|}
\multicolumn{4}{c}{\textbf{LOCAL TIMESCALE}} \\
\firsthline
parameter & intermediate lattice & starting lattice & final lattice \\
\hline
\multicolumn{4}{c}{DRIVEN SYSTEM} \\
\hline
$V_{\overline{X},X,\widetilde{Y},Z} \left(E_R\right)$ & 8.1(2),0.23(1),9.3(3),8.5(2) & \multicolumn{2}{c|}{24.1(7),2.7(1),13.4(4),11.1(3)}  \\
$t_{x,w,y,z}/h \left(\text{Hz}\right)$ & 510(40),32(1),100(6),103(6) & 200(30),0.3(03),40(3),40(3) & - \\
$t^{\text{eff}}_{x,w,y,z}/h \left(\text{Hz}\right)$ & - & - & 80(10),0.3(03),40(3),38(3) \\
$U/h \left(\text{Hz}\right)$ & 1140(20) & \multicolumn{2}{c|}{500(30)} \\
$d_x/(\lambda/2)$ &  & \multicolumn{2}{c|}{0.81(1)} \\
\hline
\multicolumn{4}{c}{UNDRIVEN SYSTEM} \\
\hline
$V_{\overline{X},X,\widetilde{Y},Z} \left(E_R\right)$ & 8.1(2),0.23(1),9.3(3),8.5(2) & 24.0(7),2.7(1),13.4(4),11.1(3) & 24.1(7),1.40(4),13.4(4),12.3(3) \\
$t_{x,w,y,z}/h \left(\text{Hz}\right)$ & 510(40),32(1),100(6),103(6) & 200(30),0.3(03),40(3),40(3) & 80(10),0.6(1),40(3),39(3) \\
$U/h \left(\text{Hz}\right)$ & 1140(20) & \multicolumn{2}{c|}{500(30)} \\
\hline 
\multicolumn{4}{c}{}\\
\multicolumn{4}{c}{\textbf{GLOBAL TIMESCALE}} \\
\multicolumn{4}{c}{DRIVEN SYSTEM} \\
\hline
$V_{\overline{X},X,\widetilde{Y},Z} \left(E_R\right)$ & 8.0(2),0.19(1),9.3(3),8.5(2) & \multicolumn{2}{c|}{23.9(7),6.1(2),9.3(3),6.0(2)} \\
$t_{x,w,y,z}/h \left(\text{Hz}\right)$ & 470(30),38(2),100(6),106(6) & 510(90),0.5(05),100(6),102(9) & - \\
$t^{\text{eff}}_{x,w,y,z}/h \left(\text{Hz}\right)$ & - & - & 210(40),0.5(05),100(6),94(8) \\
$U/h \left(\text{Hz}\right)$ & \multicolumn{3}{c|}{710(20)} \\
$d_x/(\lambda/2)$ &  &\multicolumn{2}{c|}{0.75(2)} \\
\hline
\multicolumn{4}{c}{UNDRIVEN SYSTEM} \\
\hline
$V_{\overline{X},X,\widetilde{Y},Z} \left(E_R\right)$ & 8.0(2),0.19(1),9.3(3),8.5(2) & 23.9(7),6.1(2),9.3(3),6.0(2) & 23.9(7),3.6(1),9.3(3),7.5(2) \\
$t_{x,w,y,z}/h \left(\text{Hz}\right)$ & 470(30),38(2),100(6),106(6) & 510(90),0.5(05),100(6),102(9) & 200(30),0.5(1),100(6),94(7) \\
$U/h \left(\text{Hz}\right)$ & \multicolumn{3}{c|}{700(20)} \\
\hline
\end{tabular}
\caption{Parameters for the off-resonant modulation in the intermediate, starting and final lattice configuration. 
For both the local and the global timescale we show the values of the lattice depth $V_{\overline{X},X,\widetilde{Y},Z}$ in the driven and undriven system. 
Errors in the lattice depths account for an uncertainty of the lattice calibration and an additional statistical error due to fluctuations of the lattice depth, which amounts to a total error on the lattice depth of 2.8\%.
The value and error on the tunneling rates $t_{x,w,y,z}$ result from the uncertainty of the lattice depth. 
$t_{x,w,y,z}^{\text{eff}}$ is calculated with the effective terms of the Hamiltonian given in Eq.\ref{Heff_off_res}.
The error in the on-site interaction $U$ additionally includes the uncertainty of the magnetic field and Feshbach resonance calibrations.  
The correction factor $d_x$ which is relevant for driving amplitude $K_0$ deviates from $\lambda/2$ for a non-cubic lattice and is given for the two lattice configurations we use in the driven case. 
}
	\label{Table_offRes}
\end{table*}

\begin{figure}[bt]
    \includegraphics[]{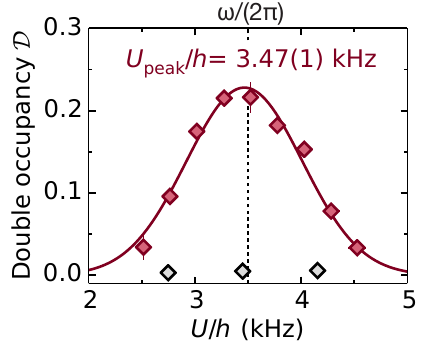}
    \caption{Response of the $\mathcal{D}$ when modulating near-resonantly at $K_0=1.43(2)$ and $\omega/(2\pi)=3.5~\text{kHz}$.   
    The solid line represents a Gaussian fit to the data to extract the peak position of the resonance. 
    The gray data points correspond to the measured value of $\mathcal{D}$ in the undriven system. 
    Data points are the mean and standard error of 3 individual measurements at different times within one driving period.
    Error bars in $U$ result from the uncertainty of the calibration. 
	}\label{SI_fig_peak}
\end{figure}

\subsection{Near-resonant modulation}

\begin{figure*}[bt]
    \includegraphics[]{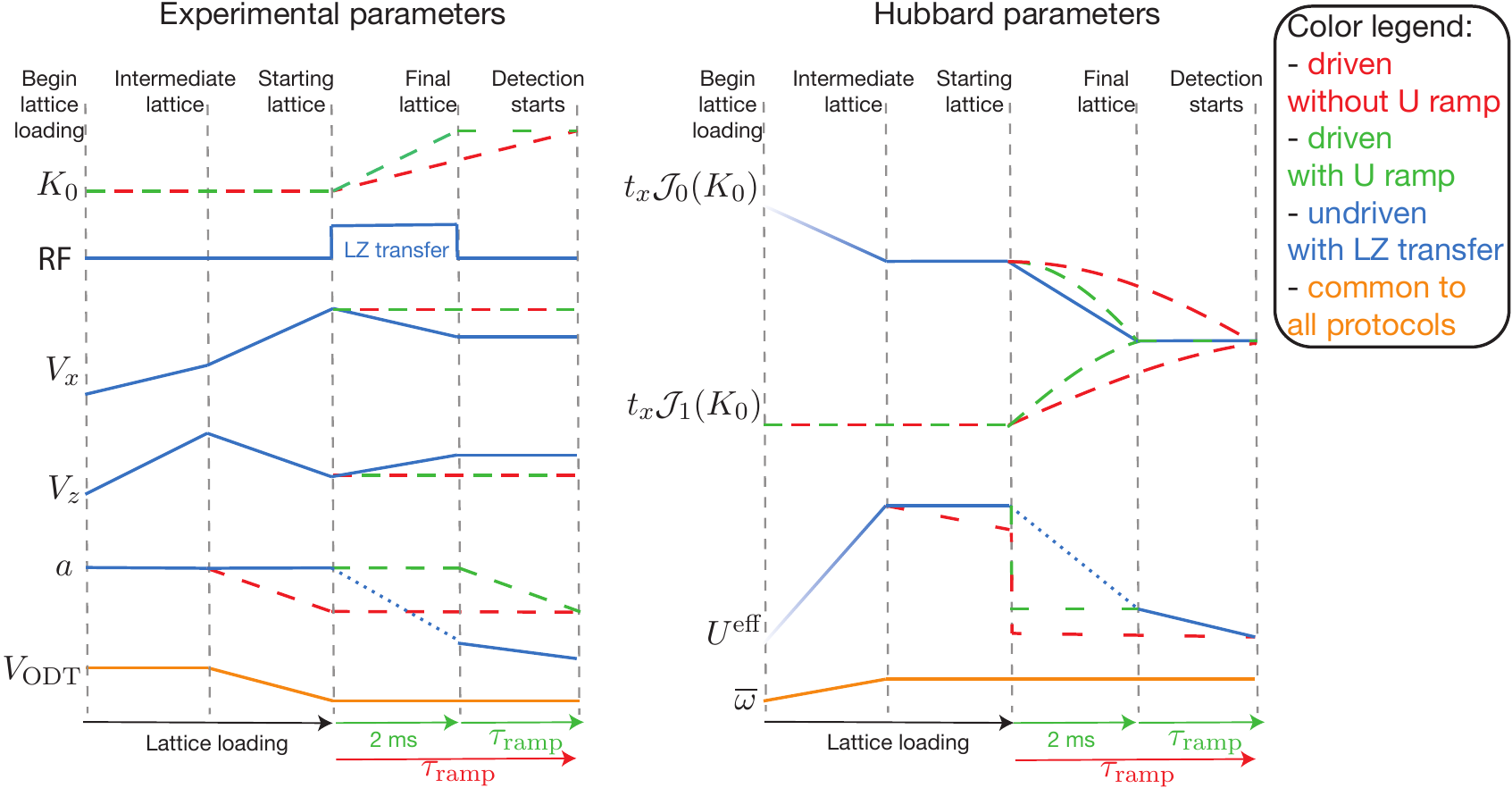}
    \caption{Ramp protocol for the near-resonantly driven Fermi-Hubbard model. The driving amplitude $K_0$, the duration of the radio-frequency pulse (RF) for the change of the interactions in the undriven case, the depth of the optical lattice $V_{x(z)}$ along the $x(z)$-direction, the scattering length $a$, and the depth of the optical dipole trap ($V_{\text{ODT}}$) are shown on the left. 
    In the driven case we use two different ramp protocols depending on the exact level of the interactions when the modulation is ramped up. 
    On the right we show the calculated parameters of the system. In the end of our protocol, the renormalized single particle tunneling $t_x \mathcal{J}_0(K_0)$, the density assisted tunneling $t_x \mathcal{J}_1(K_0)$ and tunneling in the undriven case all reach the same level. 
    A similar protocol allows to reach the same final interaction $U$ although the path in parameter space is different for the protocols.   
    We adjust $V_{\text{ODT}}$ such that the mean trapping frequency is kept constant for all lattice configurations.   
	}\label{SI_fig_NR_protocol}
\end{figure*}

In the near-resonantly driven case we start with a Mott insulator by preparing a balanced -9/2, -5/2 spin mixture with strongly repulsive interactions.
Again, we first load an intermediate lattice configuration (see Fig.~\ref{SI_fig_NR_protocol} and Table~\ref{Table_Res}) in 200~ms with $t_w/h = 37(2)\, \text{Hz}$.
During a second lattice ramp (10~ms) we load the starting lattice which has equivalent tunneling but $t_w/h < 1\, \text{Hz}$. 
We also ramp down $V_{\text{ODT}}$ to keep $\overline{\omega}$ constant.

For the driven case we use $\omega/(2\pi)=3.5~\text{kHz}$ and choose $K_0=1.43(2)$ such that the renormalized single particle tunneling $t_x \mathcal{J}_0(K_0)$ is equivalent to the density assisted tunneling $t_x \mathcal{J}_1(K_0)$.   
Depending on the effective interaction $U^{\text{eff}}= U-\hbar \omega$ we can realize a system with effective attractive or repulsive interactions. 
When driving exactly on resonance we can mimic an effective non-interacting  interaction (see Table~\ref{Table_Res}).
To calibrate the resonance peak we measure the response in $\mathcal{D}$ as a function of the interactions $U$. 
At every value of $U$ we ramp up the drive to $K_0=1.43(2)$ within 10~ms and detect the resulting $\mathcal{D}$ (see Fig.~\ref{SI_fig_peak}).
By fitting a Gaussian distribution to the data we find the peak at $U_{\text{peak}}/h=3.47(1) \, \text{kHz}$ which is close to the expected value. 
This experimentally measured value is then used when we calculate the effective interactions in the driven system ($U^{\text{eff}}=U-U_{\text{peak}}$).

We follow two different ramp protocols which allow to prepare different Floquet states \cite{Desbuquois2017}. 
Therefore, we define the loading value of $U_{\text{load}}$ as the interactions of the system when we start the drive. 
The protocols are differentiated by ramping the interactions from $U_{\text{load}}$ to $U$ during the drive or keeping it at a fixed value. 
For one ramp protocol we first ramp the drive amplitude $K_0$ from 0 to $1.43(2)$ within 2~ms at $U_{\text{load}}/h= 4.65(9)\, \text{kHz}$ detuned from the resonance.
This realizes a system with effective interactions of $\approx 1.15 \, \text{kHz}$.
At the end of this ramp the density assisted tunneling and the single particle tunneling are both renormalized to $t_x^{\text{eff}}/h=110(20) \, \text{Hz}$.
Subsequently, we tune the interactions on a variable time $\tau_{\text{ramp}}$ in the driven system, by changing the magnetic field, at constant $K_0$ ($U_{\text{load}} \rightarrow U$).
To achieve a varying effective interaction we ramp to three different values of $U$ always using the same value of $U_{\text{load}}$ (see Table~\ref{Table_Res}). 
For example, ending the ramp at $U/h=4.18(8) \, \text{kHz}$ the drive with frequency $\omega/(2\pi)=3.5 \, \text{kHz}$ leads to an effective interaction of $U^{\text{eff}}/h= 0.69(8)\, \text{kHz}$.
 
The other ramp protocol follows a different approach. 
Here, starting from $U_{\text{load}}=4.63(10)\, \text{kHz}$ we first ramp the interactions to the final value $U$ to match the desired effective interaction within 2~ms. 
Then, we ramp up the driving amplitude to $K_0=1.43(2)$ within a variable time $\tau_{\text{ramp}}$.
While driving, the static interaction $U$ is constant, however we enter the regime of effective interactions and get $U-\hbar \omega$.
To vary the effective interactions we start the drive at three different values of $U$ (see Table~\ref{Table_Res}).

To mimic the changes in an undriven system we have to change both the interactions $U$ as well as the tunneling $t_x$. 
Here, we use the fact that for the same magnetic field the Feshbach resonance of the $-9/2,-7/2$ spin mixture leads to a different scattering length $a$ as the one of the $-9/2,-5/2$. 
Starting from a strongly repulsive system in the $-9/2,-5/2$ mixture we perform a Landau-Zener transfer of the -5/2 to the -7/2 state, thereby realizing weakly repulsive interactions. 
During this transfer of 2~ms we simultaneously ramp the lattice parameters to ramp down $t_x$ to the values of the final lattice (see Fig.~\ref{SI_fig_NR_protocol} and Table~\ref{Table_Res}). 
In a final step, we ramp $U$ on a variable time by changing the scattering length to reach the final value of the interactions corresponding to $U^{\mathrm{eff}}$.
The final tunneling rates and interactions are equivalent in all three protocols which allows us to compare the evolution of $\mathcal{D}$.

Furthermore, we perform two set of measurements to obtain reference values of $\mathcal{D}$ in the starting and final lattice configurations.
While the starting lattice is in the deep Mott insulating regime and shows negligible fraction of double occupancies the final lattice configuration is prepared at $U$ that is weakly attractive, repulsive or zero. 
For weak interactions we load the  $-9/2,-7/2$ spin mixture in the final lattice configuration.
For these reference measurements we first load the system within 200~ms into an intermediate lattice with equivalent tunneling rates in $t_{x,y,z}$ as the desired lattice but $t_w\neq0$. 
During a second ramp, lasting 10~ms, we then load the final lattice by suppressing $t_w$ to a value $< 1\, \text{Hz}$.

\begin{table*}[bt]
\setlength\extrarowheight{5pt}
\begin{tabular}{|c || c | c | c|}
\multicolumn{4}{c}{\textbf{NEAR RESONANT MODULATION}} \\
\firsthline
parameter & intermediate lattice & starting lattice & final lattice \\
\hline
\multicolumn{4}{c}{DRIVEN SYSTEM} \\
\hline
$V_{\overline{X},X,\widetilde{Y},Z} \left(E_R\right)$ & 10.0(3),0.10(1),9.4(3),9.0(3) & \multicolumn{2}{c|}{24.0(7),3.7(1),9.4(3),7.3(2)} \\
$t_{x,w,y,z}/h \left(\text{Hz}\right)$ & 210(20),37(2),98(6),101(6) & 200(30),0.5(05),98(6),98(7) & - \\
$t^{\text{eff}}_{x,w,y,z}/h \left(\text{Hz}\right)$ & - & - & 110(20),0.5(05),98(6),96(7) \\
$U/h \left(\text{Hz}\right)$ & 4630(100) & [2740(50),3460(60),4180(80)] & - \\
$U^{\text{eff}}/h \left(\text{Hz}\right)$ & - & - & [-720(50), 20(60), 690(80)] \\
$d_x/(\lambda/2)$ &  &\multicolumn{2}{c|}{0.82(1)} \\
\hline
\multicolumn{4}{c}{UNDRIVEN SYSTEM} \\
\hline
$V_{\overline{X},X,\widetilde{Y},Z} \left(E_R\right)$ & 10.0(3),0.10(03),9.3(3),9.0(3) & 24.0(7),3.7(1),9.3(3),7.3(2) & 24.0(7),2.5(1),9.3(3),8.0(2) \\
$t_{x,w,y,z}/h \left(\text{Hz}\right)$ & 200(20),38(2),99(6),101(6) & 200(30),0.5(1),99(6),98(7) & 110(20),0.6(1),99(6),97(6) \\
$U/h \left(\text{Hz}\right)$ & 4800(100) & 1190(20) & [-770(50),-20(30),710(20)] \\
\hline
\end{tabular}
\caption{Parameters for the near-resonant modulation in the intermediate, starting and final lattice configuration in the driven and undriven system. 
Errors and numerical calculations as in Table~\ref{Table_offRes}. 
The effective interactions $U^{\text{eff}}$ and the effective tunneling rates $t_{x,w,y,z}^{\text{eff}}$ are calculated with the effective terms of the Hamiltonian given in Eq.2.
For $U^{\text{eff}}$ we first take a resonance curve and detect the peak of the resonance to get an experimental value for the shaking response. 
This leads to a small deviation from the expected value when using directly the shaking frequency but is within the uncertainty of $U$. 
}
	\label{Table_Res}
\end{table*}

\begin{table*}[bt]
\setlength\extrarowheight{5pt}
\begin{tabular}{|c || c | c | c|}
\multicolumn{4}{c}{\textbf{TUNABLE BANDGAP}} \\
\firsthline
parameter & intermediate lattice & starting lattice & final lattice \\
\hline
\multicolumn{4}{c}{SIMPLE CUBIC LATTICE} \\
\hline
$V_{\overline{X},X,\widetilde{Y},Z} \left(E_R\right)$ & 6.5(2),0,9.3(3),9.3(3) & \multicolumn{2}{c|}{6.5(2),0,9.3(3),9.3(3)} \\
$t_{x,w,y,z}/h \left(\text{Hz}\right)$ & 200(10),200(10),100(6),100(6) & \multicolumn{2}{c|}{200(10),200(10),100(6),100(6)} \\
$U/h \left(\text{Hz}\right)$ & \multicolumn{3}{c|}{710(10)} \\
\hline
\multicolumn{4}{c}{DIMERIZED LATTICE} \\
\hline
$V_{\overline{X},X,\widetilde{Y},Z} \left(E_R\right)$ & 10.0(3),0.11(03),9.3(3),9.0(3) & \multicolumn{2}{c|}{10.0(3),0.11(03),9.3(3),9.0(3)} \\
$t_{x,w,y,z}/h \left(\text{Hz}\right)$ & 210(20),36(2),100(6),101(6) & \multicolumn{2}{c|}{210(20),36(2),100(6),101(6)} \\
$U/h \left(\text{Hz}\right)$ & \multicolumn{3}{c|}{710(20)} \\
\hline
\multicolumn{4}{c}{HEXAGONAL LATTICE} \\
\hline
$V_{\overline{X},X,\widetilde{Y},Z} \left(E_R\right)$ & 9.9(3),0.10(03),9.3(3),9.0(3)) & \multicolumn{2}{c|}{24.2(7),3.7(1),9.3(3),7.2(2)} \\
$t_{x,w,y,z}/h \left(\text{Hz}\right)$ & 210(20),38(2),100(6),102(6) & \multicolumn{2}{c|}{180(30),0.5(05),100(6),104(7)} \\
$U/h \left(\text{Hz}\right)$ & \multicolumn{3}{c|}{710(20)} \\
\hline
\end{tabular}
\caption{Parameters for the three lattice configurations used to tune the bandgap and measure the atom loss as a function of the shaking frequency in Fig.~\ref{fig4}. 
Errors and calculations as in Table~\ref{Table_offRes}.
}
	\label{Table_bandgap}
\end{table*}


\begin{thebibliography}{63}
\bibitem [{Sha()}]{Shaking}%
  \BibitemOpen
  \href@noop {} {}\bibinfo {note} {We find the dispersion and band gaps along
  the driving direction to be most relevant for atom loss.}\BibitemShut {Stop}%
\end{thebibliography}
\end{document}